\documentstyle[preprint,aps]{revtex}
\tightenlines 
\begin{document}
\draft

\title{Polarization-dependent discharge in fibers of semiconducting ladder-type polymer}
\author{N. Kumar and K.S. Narayan \footnote{Electronic mail: narayan@jncasr.ac.in}} 

\address{Chemistry and Physics of Materials Unit, Jawaharlal Nehru Centre
for Advanced Scientific Research, Jakkur P.O., Bangalore 560 064, India.}
\date{\today}

\maketitle

\begin{abstract}
We report results on polarization-dependent photoinduced discharge in oriented fibers and films of ladder-type, electron-transporting polymer poly (benzimidazobenzophenanthroline), BBL. The photocarrier generation efficiency ($\phi$) in the fiber which is indicated by the rate of discharge, is found to be distinctly higher for light polarized parallel to the fiber axis (P$_\|$) as compared to the radially perpendicular direction (P$_\bot$). Similar results, with $\phi$ anisotropy $\sim$ 10 are obtained for oriented films.
These observations are different from previously obtained results on polyparaphenylenevinylene (PPV). The results are compared with the polarization-dependent steady- state photoconductivity measurements. We  interpret these results on the basis of molecular and macroscopic features of the material.
\end{abstract}

\narrowtext

The phenomenon of photoconductivity in conjugated polymer based devices has been extensively studied in recent years \cite{RefJ1,RefJ2,RefJ3,RefJ4,RefJ5,RefJ6}. Many models have been developed to explain the various processes involved in charge generation and transport. An interesting aspect, which has not yet been clearly understood, and which is rarely observed is the polarization-dependent anisotropy of photocurrent (I$_{ph}$).  Difficulties are encountered both in the measurement and in the interpretation of the polarization dependence of I$_{ph}$. This is primarily due to the fact that the photocurrent is a bulk process with the total efficiency governed by the free carrier generation and the subsequent transport of the charge carriers. The latter processes involved in I$_{ph}$ are independent of polarization of the incident light. However, intrinsic optical properties are highly anisotropic due to the asymmetric nature of the polymer structure. Absorption experiments are widely used to characterize optical dichroism in thin polymer films. The macroscopic fluorescence tensor is determined by the single molecule fluorescence tensor and the orientational transformation matrix that relates the molecular and macroscopic axes. Their elements are linked with orientational order parameters characterizing the orientational
distribution function \cite{RefB}. Hence, polarized fluorescence
spectroscopy and of lately, polarized electroluminescence, EL has been extensively used to study
orientational characteristics of anisotropic samples such as
liquid crystalline materials and stretched polymers \cite{RefJ7}.
The observation of a polarization-dependent bulk feature thus requires samples with high degree of orientation along with reduced disorder, since disorder drastically affects the optical and electrical properties, which can lead to ambiguous interpretation.

There have been conflicting reports on photoconductivity (PC) measurements on stretched PPV derivative films which have indicated both absence of any polarization-dependent features and in certain instances a substantial presence of polarization dependence \cite{RefJ1,RefJ4,RefJ5}. The presence of the I$_{ph}$ anisotropy as observed by the transient sub-nanosecond PC measurements, with higher I$_{ph}$ and longer lifetime for carriers generated by P$_\bot$, were interpreted on the basis of interchain photoexcitation processes \cite{RefJ1}. The interchain processes involve electrons and holes generated on separate chains, which have a lower probability of recombination than that of intrachain photoexcitation processes. On the other hand, the absence of I$_{ph}$ anisotropy, especially the fast component has been explained on the basis of generation of hot carriers with a high kinetic energy which is weakly dependent on the lattice temperature and chain structure which includes anisotropy in transfer integrals, chain ends and chemical defects \cite{RefJ5}. 

We address this problem using a modified xerographic technique on the ladder type polymer BBL. The conjugated ladder polymer BBL, with its extensive pi-electron delocalization has drawn much attention due to its remarkable electronic and opto-electronic properties. The polymer possesses excellent mechanical and thermal properties, which can be processed to have high degree of orientation \cite{RefJ8,RefJ9}. The fiber formation property  representing aligned polymer chains is another interesting feature in this system.  

The xerographic technique has evolved as an accurate method of determining the photogeneration efficiency of photoreceptors \cite{RefJ10,RefJ11,RefJ12,RefJ13}. The surface potential discharge  represented by the photo induced discharge curve (PIDC), is a direct measure of the photogenerated charge carriers \cite{RefJ11}. The photo induced discharge characteristics can be further separated into emission -limited and space-charge-limited cases. The absorbed photons create electron-hole pairs, which drift away from the photogeneration region, thus reducing the electric field. The rate of change of electric field is given by  dE/dt = e$\eta$AI$_{0}$/$\epsilon_{s}$  where I$_{0}$ is the flux, A is the absorption, and $\eta$ is the supply efficiency indicating the number of carriers in the photoabsorption region injected into the bulk, which in emission limited case equals the quantum efficiency for conversion of photon to free carriers. A distinct advantage of this method compared to electrode contact based I$_{ph}$ measurements is the absence of photoinjection and interfacial effects which can mask out subtle intrinsic features \cite{RefJ3}. The present method employed is capable of probing the photogenerated charge carriers at length scales equivalent to migration distances of $\sim$ 100 $\AA$. In our knowledge this technique has not been used with polarized light to study the efficiencies of charge generation in oriented polymer systems. 

The oriented film and fiber ($\sim$ 10 $\mu$m in diameter) of BBL were used for the present studies and were prepared by a spinning process \cite{RefJ8}. In the Xerographic experiment, the surface of the fiber was charged by corona and the surface potential decay with time was measured by a probe which was capacitively coupled to the back plate, the voltage being measured by a Keithley Inc. electrometer having an input resistance of 10$^{14}$ $\Omega$. The surface voltage was sampled at a rate of 50 ms. A linear polarizer was placed in conjunction with an unpolarized red LED source to obtain the desired polarization. The incident photon energy of 1.84 eV is close to the bandedge for BBL ( 1.67 ev). The intensity was independent of the polarization axis and was $\sim$ 10$^{12}$ photons/cm$^{2}$s. The light was incident normal to the plane of the fiber with the polarization direction in the plane of the fiber either parallel or perpendicular to the fiber axis.  Argon atmosphere was maintained during the discharging process to avoid moisture induced effects on the surface. The results obtained were corrected for the dark decay and anisotropy in reflection for the different angles of polarization. The minor correction factor due to the small difference in the reflectance of the samples for different polarization, was taken into account using the procedure adopted by Ref. 5.
The polarization dependant PL was carried out at 80 K by placing a linear polarizer in between the sample and the xenon source with a peak wavelength of 532 nm for excitation.  The steady-state I$_{ph}$  was measured in the planar - surface geometry with an inter electrode spacing of 100 $\mu$m using standard lock-in techniques \cite{RefJ6}. It is to be noted that in case of the I$_{ph}$ (coplanar measurements) the external applied field was in the direction of the fiber axis and the polarization axis chosen was either parallel or perpendicular to this bias field. 

SEM observations of the polymer fiber reveal aligned fibrillar bundled structures as shown in Fig. 1 with the fibril diameter $\sim$ 100 nm. These fibers show features in X-ray measurements indicating that polymer chains are aligned along the fibril axis, and correlates the macroscopic and microscopic anisotropy in this system\cite{RefJ9}.

It is found that the negative charge acceptance (surface charge density) on BBL films / fibers is much higher than positive charge acceptance. It is also observed that the derivative of initial decay of PIDC is higher ($\sim$ 3 times) when the surface was charged negative indicating electrons as the predominant charge carriers in BBL . This procedure was also applied for MEHPPV films where the derivative of initial decay of PIDC was higher when the surface was charged positive indicating  hole transport, which is well established in MEHPPV. Photo-induced surface potential decay measurements on unoriented cast films of BBL is shown in Fig. 2a. The PIDC was found to be independent of the excitation polarization direction. The PIDC for the fiber with polarization of the excitation light P$_\|$ and P$_\bot$ is shown in Fig. 3. It can be seen from the figure that the derivative of PIDC is almost twice higher for P$_\|$ than for P$_\bot$. Similar results were obtained for the oriented film as shown in Fig. 2b. The photoinduced discharge rate was measured as a function of the angle of the polarization axis and is depicted in Fig. 3.(inset). The angular response can be represented by a periodic function $\simeq$  F$\cos^{2}x$ + A, where F is a slowly varying function of x, as shown in Fig. 3(inset).

Conventional contact based steady-state I$_{ph}$ of the oriented film in the planar configuration did not vary considerably between P$_\|$ and P$_\bot$ modes.  This was in contrast to the xerographic measurements and the non-dependence of I$_{ph}$ on polarization was present over a large range of voltage bias, 0 V - 200 V. 
The present results of (I$_{ph}\|$ $>$ I$_{ph}\bot$) from PIDC measurements are quite different from anisotropy reports using sub-nanosecond transient photoconductivity measurements of oriented MEHPPV and polyacetylene films, \cite{RefJ1} where higher I$_{ph}$ and longer lifetime are observed for carriers generated by light polarized perpendicular to the orientation axis. An implicit assumption here was that I$_{ph}$ was intrinsic in nature, and free carrier generation was due to direct field induced dissociation of the bound interchain polaron pair and was inferred as a measure of the interchain to intrachain processes. \cite{RefJ1,RefJ2}. 

In the present case of BBL both intrinsic and extrinsic processes exists by which initial intrachain processes such as dissociation of intrachain excitons result with electrons and holes which are spatially separated (across the chain). The following features of BBL polymer are considered to interpret the PIDC anisotropy results : (i) The alignment of the chains along the fibril axis can result in large absorption anisotropy(absorption$_{intrachain}$/absorption$_{interchain}$). This feature arises from the transition dipole moment vector largely along the chain axis as indicated by semi-empirical calculations on this system.(ii) BBL is a hetero-aromatic ladder type polymer where the initial electron-hole recombination can be effectively reduced due to the network of multiple conjugation paths. This argument is consistent with the low intrachain radiative recombination yield obtained in BBL resulting in a practically nonexistent EL and a low PL efficiency. The low PL yield, which is largely a measure of the intrachain e-h recombination, when examined carefully with polarized light excitation, shows a higher yield for P$_\|$ compared to P$_\bot$ direction. The intrachain electron-hole pair can then dissociate in presence of electric field due to combination of factors which includes thermally accessible defect-mediated processes and interchain coupling mechanisms, which finally leads to sizable I$_{ph}$ efficiency as observed in this polymer \cite{RefJ14}.  

The results of the angular dependence of the discharge rate in Fig. 3 (inset) can then be explained on the basis of the above two factors.
However, an exact quantitative estimate of the PIDC anisotropy result is difficult due to following reasons: (i) The SEM image in Fig. 1 reveals  structures which are not aligned along the fiber axis. At a molecular level, the off-axis dipole moment can also result in intra-chain processes for P$_\bot$, thereby reducing the anisotropy magnitude. (ii) The recombination processes are expected to be different for carriers generated by P$_\|$ and P$_\bot$.   

 The absence of any appreciable polarization effect in the steady-state I$_{ph}$ measurements in coplanar geometry with a sizable presence in the xerographic measurements emphasizes the more local nature of probing of the later technique. The sizable density of electronic defects in the system leading to trap limiting processes prevails over the photo-induced charge carrier generating rates and can mask out the polarization-dependent initial charge generating features. An analysis of the defect dominated kinetics and polarization-dependent transient studies can resolve this apparent contradiction of the steady-state I$_{ph}$ results with the xerographic measurements. The effects of macroscopic factors on the dc conductivity ($\sigma$$_{dc}$) is sizable due to the barriers involved in the electronic transport between the fibrils and the inhomogeneous extrinsic impurity distribution which are more obvious from anisotropy of nearly four orders as observed in $\sigma$$_{dc}$ measurements \cite{RefJ15}. Hence, the microscopic orientation observed in the fibers consisting of nearly-aligned polymer chains along with the PDIC results is a measure of the local anisotropy,  distinguishing it from the larger length scale processes.

In summary, a distinct anisotropy in the PDIC was observed as a function of polarization. The measurements confirm the existence of polarization induced effects even when it is not observable in conventional contact based photoconductivity experiments. The results of these measurements obtained are strikingly different to earlier results on PPV based polymers in the following aspects: (I) The negative charges /electron transport in BBL compared to the hole based transport in PPV systems. (II) P$_\|$ yields more efficient discharge rates compared to P$_\bot$, and can be attributed to the absorption anisotropy, which can prevail in this system accompanied by other intrachain delocalizing processes of the e-h pair.

K.S.N thanks R.J. Spry, Polymer branch, AFRL for providing the polymer fibers and Department of Science and Technology (Indo-Israel project) for partial funding.

\begin{figure}

\caption{SEM images of oriented fibers along with the chemical structure of BBL}
\label{fig:1}
\end{figure}

\begin{figure}
\caption{(a) PDIC of cast, unoriented films of BBL, dashed line is the dark background decay (b) PDIC of oriented BBL films with incident light polarization axis $\|$ and $\bot$ to orientation direction}
\label{fig:2}
\end{figure}

\begin{figure}
\caption{(a)PDIC of oriented fiber with incident polarization axis $\|$ and $\bot$ to orientation direction. (insert) The angular dependence of the dV/dt obtained from the PDIC curve for oriented film.}
\label{fig:3}
\end{figure}

\end{document}